\documentclass[sigconf]{acmart}
\AtBeginDocument{%
  \providecommand\BibTeX{{%
    \normalfont B\kern-0.5em{\scshape i\kern-0.25em b}\kern-0.8em\TeX}}}

\copyrightyear{2022} 
\acmYear{2022} 
\setcopyright{acmcopyright}\acmConference[RecSys '22]{Sixteenth ACM Conference on Recommender Systems}{September 18--23, 2022}{Seattle, WA, USA}
\acmBooktitle{Sixteenth ACM Conference on Recommender Systems (RecSys '22), September 18--23, 2022, Seattle, WA, USA}
\acmPrice{15.00}
\acmDOI{10.1145/3523227.3546773}
\acmISBN{978-1-4503-9278-5/22/09}

\usepackage{graphics}
\usepackage{graphicx}
\usepackage{amsmath}
\usepackage{booktabs}
\usepackage{multirow}
\usepackage{xcolor}
\usepackage{bm}
\usepackage{subcaption}

\begin{document}

\title[Item-Attribute-Change-Aware Recommendation in the Growing E-commerce Environment]{CAEN: A Hierarchically Attentive Evolution Network for Item-Attribute-Change-Aware Recommendation in the Growing E-commerce Environment}

\author{Rui Ma}
\authornote{The first three authors have equal contributions to this paper.}
\affiliation{%
  \institution{Alibaba Group}
  \city{Beijing}
  \country{China}
}
\email{muyu.mr@alibaba-inc.com}
\author{Ning Liu}
\authornotemark[1]
\affiliation{%
  \institution{Tsinghua University}
  \city{Beijing}
  \country{China}
}
\email{atliuning@gmail.com}

\author{Jingsong Yuan}
\authornotemark[1]
\affiliation{%
  \institution{Alibaba Group}
  \city{Beijing}
  \country{China}
}
\email{jingsong.yjs@alibaba-inc.com}

\author{Huafeng Yang}
\affiliation{%
  \institution{Alibaba Group}
  \city{Beijing}
  \country{China}
}
\email{huafeng.yhf@alibaba-inc.com}

\author{Jiandong Zhang}
\affiliation{%
  \institution{Alibaba Group}
  \city{Beijing}
  \country{China}
}
\email{chensong.zjd@taobao.com}

\renewcommand{\shortauthors}{Rui Ma, Ning Liu, Jingsong Yuan, Huafeng Yang, and Jiandong Zhang}

\begin{abstract}
Traditional recommendation systems mainly focus on modeling user interests. However, the dynamics of recommended items caused by attribute modifications (\textit{e.g.}\ changes in prices) are also of great importance in real systems, especially in the fast-growing e-commerce environment, which may cause the users’ demands to emerge, shift and disappear.
Recent studies that make efforts on dynamic item representations treat the item attributes as side information but ignore its temporal dependency, or model the item evolution with a sequence of related users but do not consider item attributes.
In this paper, we propose \textbf{C}ore \textbf{A}ttribute \textbf{E}volution \textbf{N}etwork (CAEN), which partitions the user sequence according to the attribute value and thus models the item evolution over attribute dynamics with these users. Under this framework, we further devise a hierarchical attention mechanism that applies attribute-aware attention for user aggregation under each attribute, as well as personalized attention for activating similar users in assessing the matching degree between target user and item.
Results from the extensive experiments over actual e-commerce datasets show that our approach outperforms the state-of-art methods and achieves significant improvements on the items with rapid changes over attributes, therefore 
helping the item recommendation to adapt to the growth of the e-commerce platform.
\end{abstract}



\maketitle

\section{Introduction}

With the rapid technology development on the internet, logistics, and payments, products from different regions and industries are constantly gathered on worldwide e-commerce platforms and then served to online customers. To better match the users with satisfying products, recommendation systems that focus on Click-Through Rate (CTR) prediction play an indispensable role in these large-scale business platforms. Existing studies on CTR prediction \cite{zhou2018deep,zhou2019deep,cho2020meantime,tan2021dynamic} have intensively modeled the long and short-term, emerging or disappeared user interests, by leveraging their past sequential behaviors with advanced deep neural structures.
These state-of-the-art methods work fine under a stable recommendation environment, but can hardly be applied to the growing e-commerce environment. 

\begin{figure*}[t!]
\centering
\includegraphics[width=12cm]{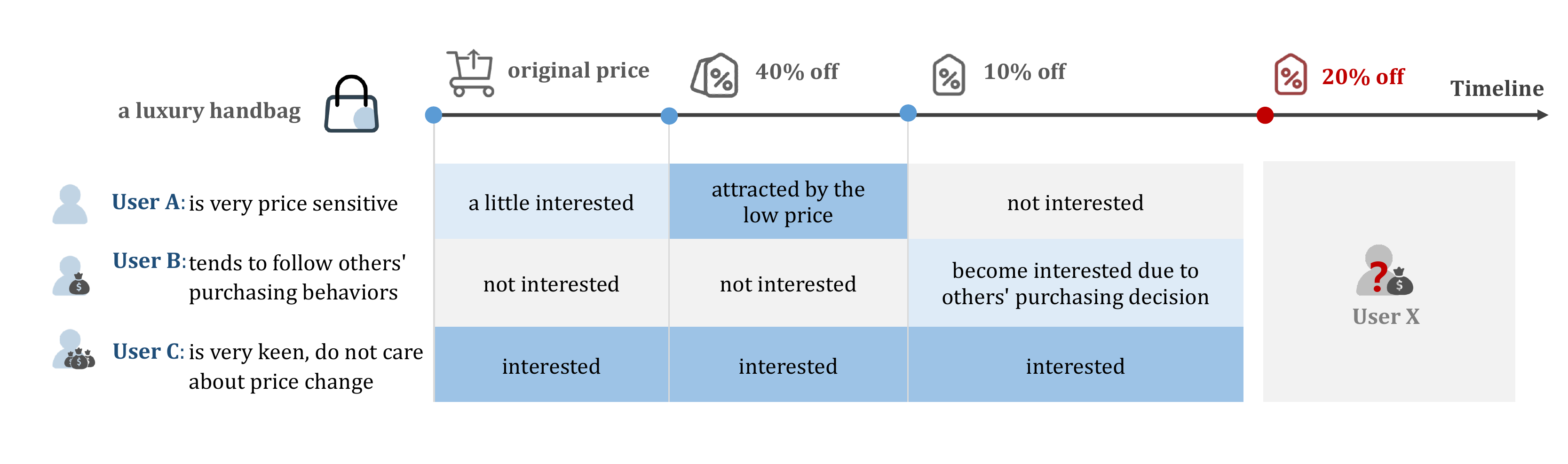}
\caption{An illustration of user preference shift caused by attribute change: Suppose that a luxury handbag is released with its original price, then gets a $40\%$ discount off during a promotion campaign, and is finally set as $10\%$ off. User A, B, C show different levels of preference for this handbag during periods with different prices.}
\label{fig:intro}
\end{figure*}

In e-commerce recommendation systems, recommended items may change rapidly due to outside interference. The interference includes both manual operations like claiming the free shipping promotion or modifying the selling price, and environmental factors like public festivals or natural disasters, which can therefore cause the users' preferences 
to emerge, shift, and disappear. For example in Figure \ref{fig:intro}, a high-priced product should have been recommended to people with high income, but once with a heavy discount, it could also meet the requirement of the price-sensitive customers. These active actions on items' core attributes from sellers can happen frequently and are encouraged by the platform operators, as they are helpful for the development of a commercial market. However, these changes could not be well captured by the structures that are based solely on users' past behaviors, which emphasizes the urgency of considering the item evolution in the growing e-commerce environment.


Recently, research efforts have been made on the dynamic item representation \cite{wu2020joint,wang2020make,wang2017topological,guo2019dynamic,li2020deep}. 
Some studies have reorganized the past user-item interaction data and make use of the users who have interacted with the target item. Specifically, \citeauthor{wang2017topological} \shortcite{wang2017topological} establish a long short-term memory network over the past interaction sequence of the item, \citeauthor{guo2019dynamic} \shortcite{guo2019dynamic} apply an attention network to extract the most similar users from the item interaction history according to the target user, and \citeauthor{li2020deep} \shortcite{li2020deep} further propose a time-aware evolution layer, which can capture the rapid item changes in popularity. 
\citeauthor{li2020deep}'s work is typical for capturing the temporal changes of items as well as modeling the dynamic item representation. However, there remain two major challenges: 
\emph{A)} The change of item attributes, which greatly affects users' decision-making outside their usual interests, should not be ignored. 
\emph{B)} Modeling item evolution under interaction granularity (regarding one interaction between user and item as one state) results in information redundancy and unnecessary complexity. Applying recurrent networks upon highly similar interactive logs leads to poor computing efficiency, thus restricting the model capacity over the time span of item history. 

In this paper, we propose \textbf{\underline{C}}ore \textbf{\underline{A}}ttributes \textbf{\underline{E}}volution \textbf{\underline{N}}etwork (CAEN) to address the aforementioned challenges. The item behavior sequence, which is a set of users in chronological order that have interacted with the target item, is partitioned into several states according to the value of its item attribute. Then, the item evolution is modeled under such attribute dynamics with corresponding users. 
Under this framework, we further devise a hierarchical attention network that includes a two-stage attentive mechanism.
The later stage applies a personalized attention layer, which activates similar users from interaction history in assessing the matching degree of the target user. More significantly, in the first stage, we apply an attribute-aware attention layer, which can accurately acquire the item representation of a certain time by putting more weights on the users that are attracted by the attribute of that time. Given that there is not yet a public dataset that includes attribute changes of recommended items, we conduct our experiment on an actual e-commerce dataset in Southeast Asia.
Experiment results show that our proposed method can effectively improve recommendation performance, especially for the items with rapidly changing attributes. The main contributions of this work can be summarized below:
\begin{itemize}
    \item To the best of our knowledge, this is the first work to exploit item attribute evolution in CTR prediction. By capturing the item change over attribute dynamics, such a framework enables the personalized item recommendations to adapt to the frequent item attribute modification in the growing e-commerce environment.
    \item We propose a hierarchical attention mechanism inside item evolution modeling. Underneath the personalized attention, attribute-aware attention is applied to improve the item representation learning by activating the interactions that are caused by the specific attribute. This mechanism is of particular benefit to our concept of attribute evolution modeling.
    \item We evaluate the proposed method on a CTR prediction task over actual e-commerce datasets in Southeast Asia. Experiment results show that our method outperforms the state of the arts, especially on the items with rapid changes over attributes, demonstrating the effectiveness of capturing the dynamics in item recommendation.
\end{itemize}

\section{Related Work}
Previous studies on personalized recommendation propose to leverage the past interactions between users and items. These methods mostly focus on modeling users' interests, while only a few consider the item attributes and dynamics.

\subsection{Sequential Modeling on User Behavior}

Early methods are mostly based on matrix factorization (MF) and Markov chains (MC) \cite{rendle2010factorizing,cheng2013you}, 
however, these methods only consider the local patterns of adjacent behaviors.
Recently, deep neural networks have attracted researchers' attention due to their advanced capabilities in sequence modeling, especially the recurrent neural networks (RNN) \cite{hidasi2015session,chen2018sequential}. 
Further, 
parallel models, including convolutional neural network \cite{tang2018personalized} and attention network \cite{zhou2018atrank,kang2018self,sun2019bert4rec}, are proposed with better efficiency in dealing with long sequences. 
In addition, the graph neural network \cite{yu2020tagnn} is also applied in sequential recommendation thanks to its excellent representation capability on structured data. 

The above methods achieve satisfactory results over corresponding datasets, which are collected in relatively stable environments. However, in the growing e-commerce environment, recommended items can change rapidly and further affect users' preferences. These methods can be incompetent when dealing with ubiquitous changes.

\subsection{Modeling on Item Attributes and Dynamics}
Recent work for item modeling in recommendations mainly includes two aspects: \textit{interaction behaviors} and \textit{item attribute} modeling.
On the one hand, rich user-item interaction information can help to capture the item dynamics in a temporal evolution. Some studies \cite{wang2017topological,guo2019dynamic,li2020deep} aims to extract dynamic representations of the item by modeling the sequential users who interact with it. 
On the other hand, item attributes are high-availability and high-value information for the recommendation system, thus they are widely used to mitigate the sparse interaction data problem. 
Recently, item attributes have been utilized as side information in both matrix factorization (MF) \cite{yu2017attributes} and deep neural networks \cite{sun2021attribute}. Moreover, some studies \cite{yuan2021icai,wang2020make} exploit the correlation between an item and its attribute in user behavior sequence.
Besides, \citeauthor{wu2020joint} \shortcite{wu2020joint} developed a multi-task framework to realize co-optimization of attribute inference and item recommendation for item embedding learning.

However, the above methods do not consider the attribute changes of a particular item and the corresponding impact on the user. Therefore, we propose to model item evolution by capturing the dynamics of item attributes and leveraging the attributes to obtain a better representation in interactions behaviors modeling.


\section{Methodology}

In this section, we first specify the research problem and the preliminary. Then we present our whole framework of item attribute evolution network. Later, we expand how we organize and model the attribute-based state, and elucidate how we emphasize the impact of item attributes on user-item interactions through a hierarchical attention network. At the end of this section, we introduce the deployment of our method into downstream applications.

\subsection{Problem Statement}
\begin{figure*}[t!]
\centering
\includegraphics[width=15cm]{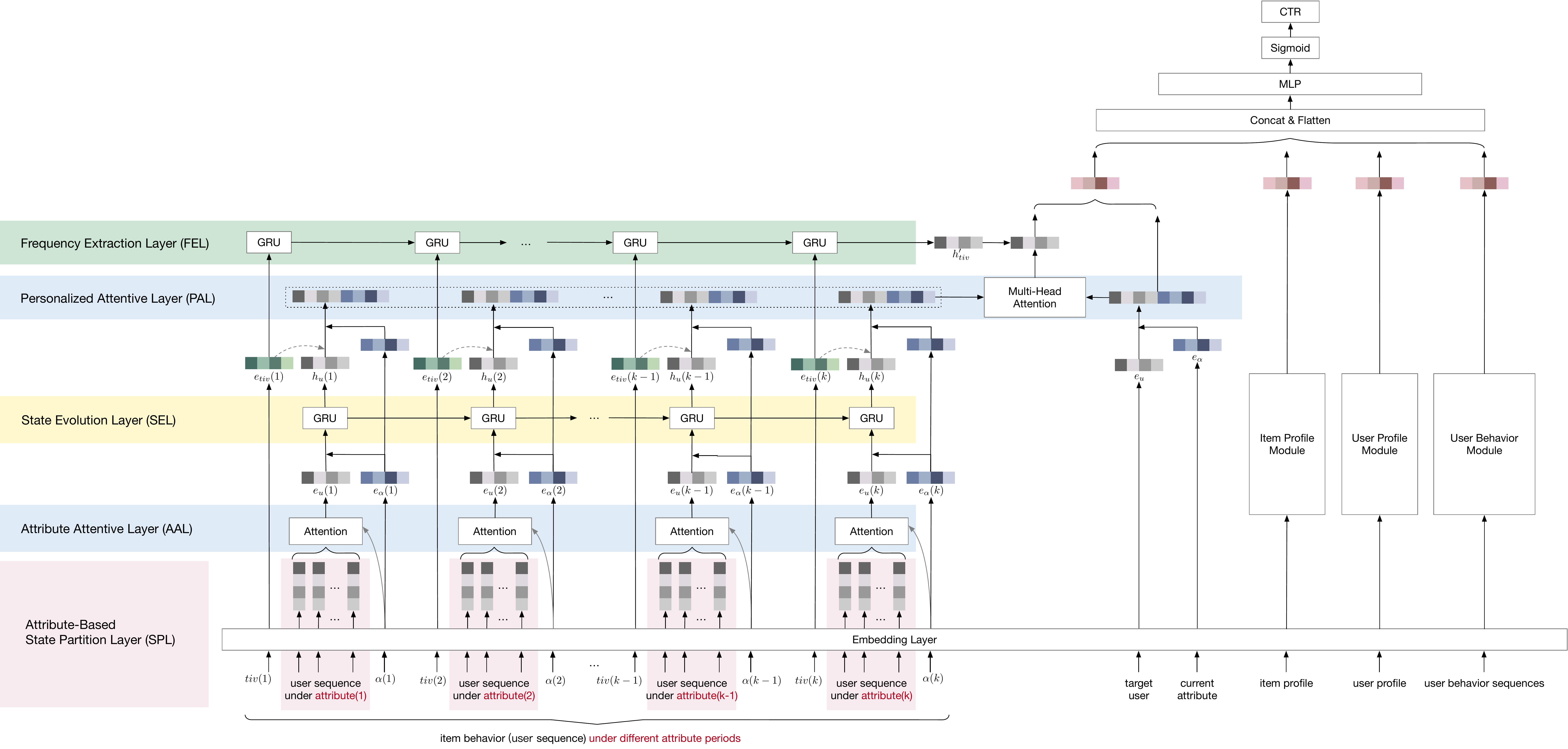}
\caption{The architecture of Core Attribute Evolution Network is presented, which corresponds to the Item Behavior Modeling Module on the right. The timeline of the item's lifecycle is partitioned into periods according to its price. Each period corresponds to an attribute state in the above architecture and is further transmitted to subsequent modeling.}
\label{fig:CAEN}
\end{figure*}

In recommendation systems, the click-through rate $y$, which is also the probability of interaction between the target user $u$ and target item $i$, is usually estimated by the information from four modules including the \textit{User Profile} $\mathcal{G}_1(u)$, \textit{Item Profile} $\mathcal{G}_2(i)$, \textit{User Behavior Modeling} $\mathcal{G}_3(u)$, and the \textit{Item Behavior Modeling} $\mathcal{G}_4(i)$, as shown on the right side of Figure \ref{fig:CAEN}. For target user $u$ and target item $i$, let $\mathcal{I}_u = \big\{i_u^{1}, i_u^{2}, ..., i_u^{N}\big\}$ denote the $N$ items that have interacted with user $u$, and $\mathcal{U}_i = \big\{u_i^{1}, u_i^{2}, ..., u_i^{M}\big\}$ denote the $M$ users that have interacted with item $i$. Therefore, the probability of interaction between $u$ and $i$ can be expressed as:
\begin{equation}\label{eq:ctr}
    \hat{y} = P_{ctr}(u,i) = \mathcal{F}\Big(\mathcal{G}_1(u), \mathcal{G}_2(i), \mathcal{G}_3(\mathcal{I}_u), \mathcal{G}_4(\mathcal{U}_i)\Big), 
\end{equation}
where $\mathcal{F}$ denotes the top decision layer.

In this paper, we mainly focus on the module of \textit{Item Behavior Modeling} $\mathcal{G}_4(i)$, which targets to leverage $\mathcal{U}_i$ in an effective way to adapt to the item evolution over its attribute dynamics.

\subsection{Core Attributes Evolution Framework}

There are serveral attributes over a given item on the e-commerce platform, among which some attributes can significantly affect the users' behavior, and can also vary over time due to different reasons. These attributes are called the core attributes, which include price, promotion, free shipping, warranty service, etc. 

Since the items change along with their attributes, we propose to partition and model the lifecycle of these items based on their attribute variations.
The framework of our CAEN is composed of $5$ layers, $2$ of which form the Hierarchical Attention Network (HAN) and will be mathematically expanded in the next subsection:
\begin{itemize}
    \item \textit{Attribute-Based State Partition Layer (SPL)}: In this layer, the lifecycle of the target item is partitioned into several states according to the change of its attribute value. A group of features under each state, i.e., the attribute value, users that have interacted with this item, as well as the time stamps under each state are organized to be embedded and fed into the next layer.
    \item \textit{Attribute Attention Layer (AAL)}: This layer is applied in each state for aggregating the related users. Instead of a simple pooling structure or a multiple pooling strategy, an attentive pooling using attribute is applied to attach importance to the interactions that are attracted by the attribute, other than the constant user interests. Therefore, this layer outputs the representations of user groups under each attribute value, which can also be seen as the representations of each attribute state.
    \item \textit{State Evolution Layer (SEL)}: Given that the item attributes are changed consecutively over time, item popularity is not only determined by its current attribute but also by its past. Here, GRU cells are applied over each attribute state to model the evolution of the target item. The output of AAL is input into a GRU cell, then the hidden state is regarded as the item representation under the corresponding attribute, and also passed to the next state.
    \item \textit{Personalized Attention Layer (PAL)}: In this layer, the target user and current attribute are concatenated as query, while the item representations are concatenated with corresponding attributes as key. Therefore, the weights of attribute states are calculated by considering both the similarity between the target user and past users, as well as the similarity between current attributes and past attributes. 
    \item \textit{Frequency Extraction Layer (FEL)}: In addition to the personalized attention, this layer aims to capture the frequency of item attribute change, which quantifies the activity level of the seller's operation, and also indicates how much of a difference this core attribute evolution network will make. 
\end{itemize}

\subsection{Modeling over Attribute-Based State: Partition, Evolution, and Frequency Extraction}
The layers of \textit{SPL}, \textit{SEL}, and \textit{FEL}, work together for the attribute-based state modeling. Among them, \textit{SEL} is a commonly used layer, which can deal with sequences at any granularity. For example, most previous methods utilizing user behavior or item behavior regard one user-item interaction as one state. However, in our method, \textit{SPL} is first applied to partition the lifecycle of the target item based on the actions from sellers and platform operators on item core attributes. Therefore, it defines the granularity of the state that \textit{SEL} works on. In \textit{SPL}, the criterion for state partition can be customized according to the purposes and characteristics of the recommendation scenario. Here, for the sake of simplicity and interpretability, we take \textit{price} as the representative in the following description.

In \textit{SPL}, the users that have interacted with the target item $i$ can be expressed as $\mathcal{U}_i = \mathcal{U}_{i}^{(1)}+ \mathcal{U}_{i}^{(2)}+ ...+ \mathcal{U}_{i}^{(k)}$, where $\mathcal{U}_{i}^{(\kappa)}$ denotes those who interacted under the $\kappa^{\rm th}$ price state. Instead of regarding one user-item interaction as one state, a group of users $\mathcal{U}_{i}^{(\kappa)}$ are seen as one state of the item. Then after \textit{AAL}, \textit{SEL} models the evolution over user groups, which is also the dynamics of the item attribute. Besides, \textit{FEL} extracts the frequency of item attribute change.  Specifically, we deploy a recurrent network over the sequence composed of the time stamps of attribute changes.

\subsection{Two-Stage Hierarchical Attention Network: User-Item Interaction under Specific Attributes}
Inspired by a series of work \cite{vaswani2017attention,yang2016hierarchical,ying2018sequential}, we build a hierarchical attention network throughout the item evolution modeling, which is a two-stage structure consisting of the \emph{Attribute Attention Layer} as well as the \emph{Personalized Attention Layer}, as mentioned in Section 3.2. In this subsection, we first introduce the mathematics of multi-head attention, then elaborate on how we implement it into our hierarchical structure.

\subsubsection{Multi-Head Attention Mechanism}
We adopt scaled the dot-product attention in the basic form of attention. With input including queries $Q$ and keys $K$ of dimension $d_k$, and values $V$ of dimension $d_v$, we can compute and scale the dot products of the query with all keys, then obtain the weights on values with a softmax function:
\begin{equation}
    {\rm Attention}(Q,K,V) = {\rm softmax}\big(\frac{QK^T}{\sqrt{d_k}}\big) V.
\end{equation}
Furthermore, linear projections are optionally applied on $Q$, $K$, $V$ to achieve higher representation capacities,  and multiple heads with different linear projections are applied to attend to information from different representation subspaces at different positions:
\begin{equation}
\begin{aligned}
    {\rm MultiHead}(Q,K,V) & = {\rm Concat}({\rm head}_1,...,{\rm head}_h)W^O \\
    {\rm head}_i & = {\rm Attention}(QW_i^Q, KW_i^K,VW_i^V), 
\end{aligned}
\end{equation}
where the projections are parameter matrices $W_i^Q\in\mathbb{R}^{d\times d_k}$, $W_i^K\in\mathbb{R}^{d\times d_k}$, $W_i^V\in\mathbb{R}^{d\times d_v}$, and $W^O\in\mathbb{R}^{h d_v\times d}$. Here $d$ denotes the model dimension.

\subsubsection{Hierarchical Attention Implementation} In Figure \ref{fig:HAN}, the implementations of the two stages of our hierarchical attention mechanism are presented respectively: 

\begin{figure}[t!]
\centering
\includegraphics[width=8.3cm]{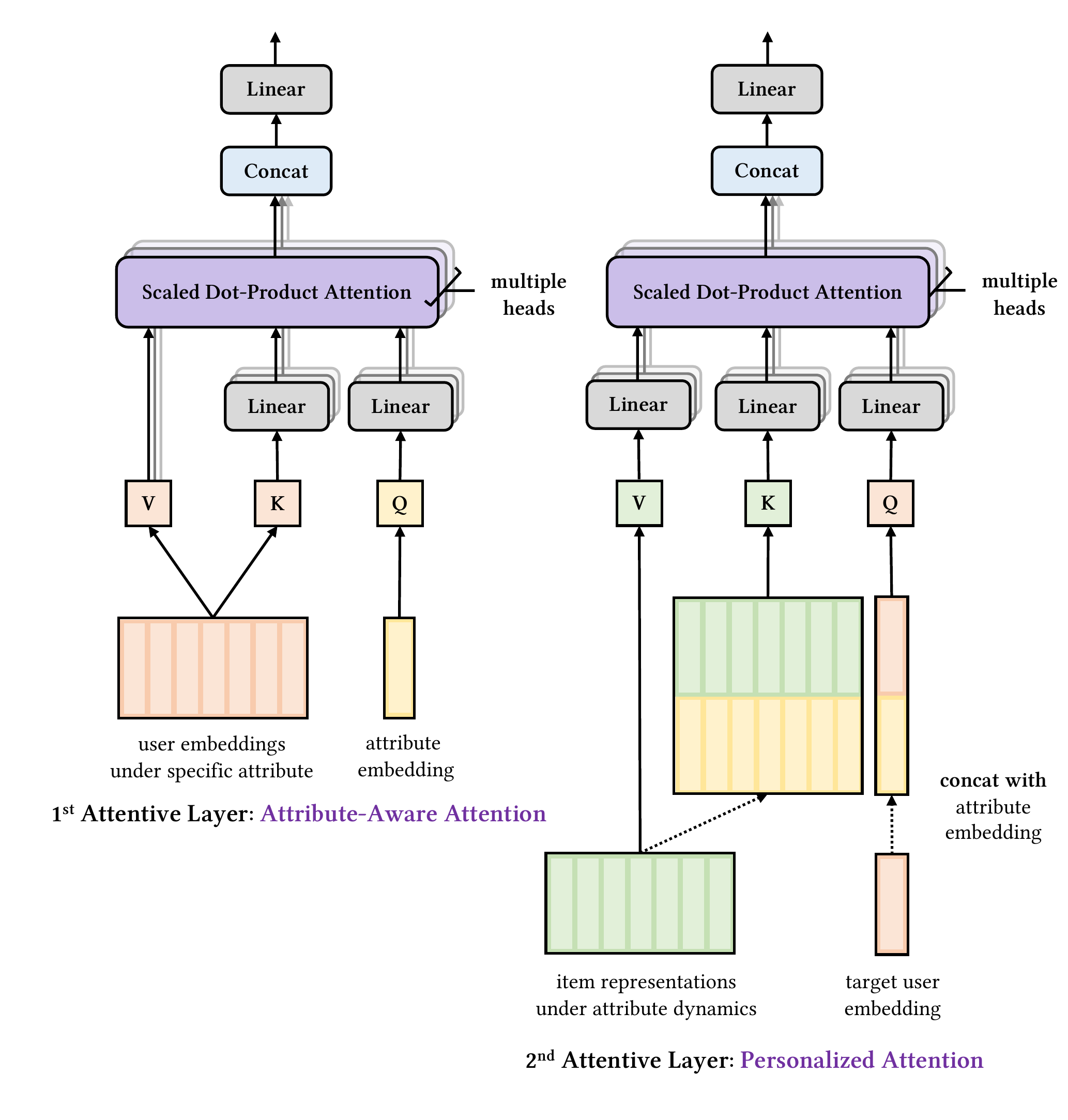}
\caption{Hierarchical Attention Network includes two attentive layers corresponding to the Attribute Attention Layer (AAL) and the Personalized Attention Layer (PAL).}
\label{fig:HAN}
\end{figure}
 

The $1^{\rm st}$ attentive layer corresponds to \textit{Attribute Attention Layer (AAL)} in CAEN, which is applied on the bottom to aggregate the users under a specific attribute. Instead of simply applying average pooling over these users, an attention mechanism is used to find the users who are most representative of the attribute among the interaction history, which is, to put more weight on the users that are attracted by the attribute, but not the users that are consistently interested in this item. To achieve this, this layer is implemented as shown on the left of Figure \ref{fig:HAN}. Query is the attribute embedding, value and key are both the embedding of the users who have interacted with the target item under this attribute:
\begin{equation}\nonumber
\begin{aligned}
{\rm Query} &= \bm{\alpha}_{\kappa}, \\ 
{\rm Key} &= U_{\alpha_\kappa}, \\ 
{\rm Value} &= U_{\alpha_\kappa},
\end{aligned}
\end{equation}
where $\bm{\alpha}_{\kappa}$ denotes the embedding of the $\kappa^{\rm {th}}$ attribute, and $U_{\alpha_\kappa}$ denotes the embedding matrix of the user sequences under the $\kappa^{\rm {th}}$ attribute:
\begin{equation}
U_{\alpha_\kappa} =
   (
        \begin{array}{cccc}
            \bm{u}^1_{\alpha_\kappa}, 
            & \bm{u}^2_{\alpha_\kappa}, 
            & ..., 
            &\bm{u}^{M_{\alpha_\kappa}}_{\alpha_\kappa}
        \end{array}
   ).
\nonumber
\end{equation}
The query and key are both projected into other spaces to obtain the appropriate similarity as attention weights. However, we keep value in its original representation space so that the output remains in the consistent space with user embedding, and thus can be compared with the target user in the following computations. Each attribute-aware attention works individually under each attribute $\alpha_1, \alpha_2, ..., \alpha_k$, but shares the same parameter $W^Q, W^k$ throughout this scheme.

The $2^{\rm nd}$ attentive layer corresponds to \textit{Personalized Attention Layer (PAL)} in CAEN, which is applied on the top to extract information from the source past interactions according to the target one. To balance the impact from attribute and the impact from user in calculating the matching degree of past interactions on current one, attribute and user are combined to act as query and key. As shown on the right, value is the item representations under different attributes that are calculated from previous layers, key is the concatenation of the item representations and the attributes, and query is the concatenation of the target user and the current attribute:
\begin{equation}\nonumber
\begin{aligned}
{\rm Query} & = {\rm Concat}\big(\bm{u}_t, \bm{\alpha}_t\big), \\
{\rm Key} & = {\rm Concat}\big(H_{\bm{\alpha}}, A \big),\\
{\rm Value} & = H_{\bm{\alpha}},
\end{aligned}
\end{equation}
where the $\bm{u}_t$ denotes the embedding of the target user, $\bm{\alpha}_t$ denotes the embedding of the current attribute, $A$ is the embedding matrix of past attributes
\begin{equation}
A =
   (
        \begin{array}{cccc}
            \bm{\alpha}_{1}, 
            & \bm{\alpha}_{2}, 
            & ..., 
            &\bm{\alpha}_{k}
        \end{array}
   ),
\end{equation}
and the $H_{\bm{\alpha}}$ is the matrix of item representations under the attribute sequence
\begin{equation}
H_{\bm{\alpha}} =
   (
        \begin{array}{cccc}
            \bm{h}_{\alpha_1}, 
            &\bm{h}_{\alpha_2}, 
            & ..., 
            &\bm{h}_{\alpha_k}
        \end{array}
   ),
\end{equation}
in which the representations are obtained from the previous \emph{State Evolution Layer}.

\subsection{Downstream Application Deployment}
The item representation from our proposed CAEN can be integrated with various downstream application structures. Here, we take CTR prediction as the downstream task and apply a Multiple Layer Perceptron (MLP) to predict the probability of the target user clicking on the candidate item, as shown on the right side in Figure \ref{fig:CAEN}.
ReLu activation is used within these layers and sigmoid activation is used at the top.
Finally, the loss function is defined as the cross-entropy for CTR prediction task, which can be formulated as follow:
\begin{equation}
    \mathcal{L}_{ctr} = -\sum_{k=1}^{N_s} [y_k \log \hat{y}_k + (1-y_k)\log(1-\hat{y}_k)],
\end{equation}
where $N_s$ is the sample size of the training dataset, $y\in \{0, 1\}$ is the label whether the user clicked the item and $\hat{y}_k$ represents the predicted probability of the user clicking on the item.

\section{Experiments}
In this section, we introduce how we set up our experiments over the real-world dataset, and answer the following four research questions based on the experiment results:
\begin{itemize}
    \item \textbf{RQ1}: How does CAEN perform compared with the state-of-the-art methods for sequential recommendations?
    \item \textbf{RQ2}: How do different components in CAEN contribute to the recommendation performance?
    \item \textbf{RQ3}: How does CAEN perform along with different methods based on user behaviors?
    \item \textbf{RQ4}: How does CAEN perform on items with different levels of attribute dynamics compared to other baselines?
    \item \textbf{RQ5}: How do the key hyper-parameters in CAEN affect its performance?
\end{itemize}

\subsection{Experimental Setups}

\subsubsection{Datasets.}
Since existing public datasets lack accurate records on item attributes, we collect a dataset from Lazada\footnote{https://www.lazada.com/}, which is a fast-growing e-commerce platform in Southeast Asia. The dataset is obtained from 8-day item exposure logs from April 18, 2022 to April 25, 2022 in Indonesia, with labels that record whether the items are clicked or not. The dataset is downsampled, and negative sampled with a ratio of 1:5.
The samples of the first seven days are used for training, and the samples of the last day are used for testing. In addition to these exposure logs, we record item attribute changes and item behaviors, \emph{i.e.}, user interactions, within 30 days before exposure. In our experiment, we choose price as attribute, then prepare relative features including discount rate, price level, and price ranking in the same category.
The statistics of the processed datasets (after downsampling) are shown in Table \ref{tab:dataset_details}.

\begin{table}[h!]
    \caption{The statistics of experimental datasets.}
    \centering
    \begin{tabular}{cc}
        \toprule
        Description & Value \\
        \midrule
        \#users & 3.4 million \\
        \#items & 3.1 million \\
        \#samples & 363 million \\
        \#category & 4,056 \\
        avg. length of the attribute states & 2.73 \\
        \bottomrule
    \end{tabular}
    \label{tab:dataset_details}
\end{table}

\subsubsection{Baselines.}
We consider three groups of competitive baselines  for performance comparison as below. 

\setlength{\parindent}{0pt}
\underline{GROUP ${\rm \uppercase\expandafter{\romannumeral1}}$} includes three non-sequential methods based on user-item interactions:
    \begin{itemize}
        \item \textbf{SVD++} \cite{koren2008factorization} treats user behaviors as a matrix, combining the neighborhood model as well as the latent factor models.
        \item \textbf{YoutubeNet} \cite{covington2016deep} proposes to obtain user representations by averaging the embedding of items in users' past behaviors.
        \item \textbf{PNN} \cite{qu2016product} introduces a product layer upon YoutubeNet to capture the similarities between candidate items and the items in past behaviors.
    \end{itemize}

    \underline{GROUP ${\rm \uppercase\expandafter{\romannumeral2}}$} includes five sequential modeling methods on only user behaviors:
    \begin{itemize}
        \item \textbf{GRU4Rec} \cite{hidasi2015session} adopts a gated recurrent unit for sequential user behavior modeling, which is known as one of the earliest works to introduce the RNN-based model for sequential recommendation.
        \item \textbf{Caser} \cite{tang2018personalized} is a CNN-based model and develops horizontal and vertical convolutional filters to capture the behavior patterns.
        \item \textbf{ATRANK} \cite{zhou2018atrank} is an attention-based model that can utilize the correlation between users’ heterogeneous behaviors.
        \item \textbf{DIEN} \cite{zhou2018deep} proposes an AUGRU to model the interest evolving process and activate specific interests for different target items, which integrates the RNN-based model and attention-based model.
        \item \textbf{UB-GRUA} \cite{li2020deep} first applies GRU in user interest modeling and then uses multi-head attention to capture the interests most relevant to the target item.
    \end{itemize}

\underline{GROUP ${\rm \uppercase\expandafter{\romannumeral3}}$} includes four dual behavior models that have both user behavior modeling and item behavior modeling. For fairness, we adopt the same user behavior modeling method in all these models as well as CAEN.
\begin{itemize}
    \item \textbf{Topo-LSTM} \cite{wang2017topological} develops LSTM to capture item dynamics from item behavior sequences.
    \item \textbf{DIB} \cite{guo2019dynamic} is an attention-like model, in which a dynamic item block searches for users who are similar to the target user from the item behaviors.
    \item \textbf{IB-GRUA} uses GRU to model the temporal dependencies among item behaviors, and an attention mechanism to generate the personalized recommendation.
    \item \textbf{TIEN} \cite{li2020deep} further introduces a time-aware structure over IB-GRUA. 
\end{itemize}

\subsubsection{Evaluation Metrics.}
We adopt two metrics to evaluate our proposed model from different perspectives, which are widely used in CTR prediction tasks. The first metric is \textbf{AUC} (Area Under ROC Curve), which indicates the pairwise ranking ability. 
Another metric is \textbf{Logloss}, also known as cross-entropy loss, which measures the likelihood of the output probability distribution and the true probability distribution. A larger AUC indicates better recommendation performance, but Logloss performs the opposite.

\subsubsection{Implementation Details.}
We implement all the models in Tensorflow and the implementation codes are released \footnote{The code is available at https://github.com/maruiRec/caen\_22}. 
The embedding dimension of item ID and user ID is set as 32 for all methods. Besides, the embedding dimension of each item attribute and user profile is set as 32 and 16 respectively. The length of the user's behavior sequence is truncated to 20, and the number of hidden units in the decision layer is fixed as [1024, 512, 128] for all models.
The code of all baselines is from the authors or open resources. And these models are all optimized by Adam optimizer with cosine annealing, where the initial learning rate is set as $1\times e^{-4}$ and the minimum learning rate is set as $5\times e^{-5}$. The training batch size is 1024.  
For the conventional item behavior modeling method, the truncation length of the interaction behavior is set to 30. For our proposed model, the truncation length of the attribute states is 8, and the truncation length of the interactive behavior under each attribute is 50. 

\subsection{Overall Performance (RQ1)}
Table \ref{tab:overall_perf} summarizes the overall performance of baselines and our method. In Group \uppercase\expandafter{\romannumeral3}, we use UB-GRUA for user behavior modeling, which has been proved to be the best model in the second group. 
According to the results, the models in Group \uppercase\expandafter{\romannumeral2} outperform most models in Group \uppercase\expandafter{\romannumeral1} with higher AUC and lower Logloss, which demonstrates the effectiveness of the state-of-the-art user behavior modeling.
However, when further establishing an item behavior modeling module, existing methods in Group \uppercase\expandafter{\romannumeral3} do not always obtain better results. Among these dual behavior models, Topo-LSTM establishes an evolution network over the past interaction sequence of the item; DIB applies an attention network to extract the most similar users from a set of users in the item interaction history according to the target user; IB-GRUA can be seen as the combination of Topo-LSTM and DIB, which considers both the chronological order of past users and their similarity to the target user. 
These three methods are proposed with good results over datasets like Amazon\footnote{http://jmcauley.ucsd.edu/data/amazon/} and MovieLens\footnote{ https://grouplens.org/datasets/movielens/}, which are collected in a relatively stable environment. However, they fail to be profitable in our fast-growing circumstance, with worse results than UB-GRUA. 

\begin{table*}[h!]
    \caption{Overall performance of our method and baseline models on AUC and Logloss metrics. FMCG, FA, GM, EL represents four industry categories of items (fast-moving consumer goods, fashion, general merchandise, and electronics). The best results within groups are with underlines, the best results of all methods are in bold.}
    \centering
    \setlength{\tabcolsep}{1.3mm}{
    \begin{tabular}{c|c|cc|cc|cc|cc|cc}
        \toprule
        \multirow{2}{*} {Group} & \multirow{2}{*}{Method}
        & \multicolumn{2}{c|} {overall} 
        & \multicolumn{2}{c|} {FMCG}
        & \multicolumn{2}{c|} {FA} 
        & \multicolumn{2}{c|} {GM}
        & \multicolumn{2}{c} {EL}\\ 
        & & AUC & Logloss & AUC & Logloss & AUC & Logloss & AUC & Logloss & AUC & Logloss\\
        \midrule
        \multirow{3}{*}{\uppercase\expandafter{\romannumeral1}}  & SVD++ & 0.6539 & 0.6689 & 0.6507 & 0.6412
         & 0.6565 & 0.6764 & 0.6474 & 0.6716 & 0.6487 & 0.6657\\
         & YoutubeNet & 0.6558 & 0.6746 & 0.6545 & 0.6447
         & 0.6584 & 0.6804 & 0.6483 & 0.6825 & 0.6491 & 0.6775\\
         & PNN & \underline{0.6564} & \underline{0.6650} & \underline{0.6552} & \underline{0.6351} 
         & \underline{0.6590} & \underline{0.6726} & \underline{0.6486} & \underline{0.6688} & \underline{0.6496} & \underline{0.6640}\\
        \midrule
        \multirow{5}{*}{\uppercase\expandafter{\romannumeral2}}
        & GRU4Rec & 0.6579 & 0.6771 & 0.6565 & 0.6457 & 0.6607 & 0.6842 & 0.6500 & 0.6829 & 0.6511 & 0.6789 \\
         & Caser & 0.6593 & 0.6708 & 0.6580 & 0.6415 & 0.6623 & 0.6768 & 0.6509 & 0.6779 & 0.6519 & 0.6730 \\
        & ATRANK & 0.6565 & 0.6567 & 0.6556 & \underline{0.6271} & 0.6591 & 0.6630 & 0.6489 & 0.6625 & 0.6499 & 0.6614 \\
        & DIEN & 0.6584 & 0.6606 & 0.6570 & 0.6320 & 0.6611 & 0.6679 & 0.6507 & 0.6644 & 0.6521 & 0.6588 \\
        & UB-GRUA & \underline{0.6620} & \underline{0.6567} & \underline{0.6618} & 0.6310 & \underline{0.6645} & \underline{0.6623} & \underline{0.6539} & \underline{0.6623} & \underline{0.6558} & \underline{0.6564} \\
        \midrule
        \multirow{4}{*}{\uppercase\expandafter{\romannumeral3}} 
        & Topo-LSTM & 0.6586 & 0.6614 & 0.6574 & 0.6303 & 0.6616 & 0.6688 & 0.6500 & 0.6671 & 0.6511 & 0.6578\\
        & DIB & 0.6590 & 0.6685 & 0.6581 & 0.6383 & 0.6618 & 0.6745 & 0.6507 & 0.6765 & 0.6519 & 0.6689\\
        & IB-GRUA & 0.6587 &\underline{ 0.6571} & 0.6580 & 0.6291 & 0.6613 & \underline{0.6625} & 0.6510 & \underline{0.6644} & 0.6521 & 0.6595\\
        & TIEN & \underline{0.6622} & 0.6591 & \underline{0.6616} & \underline{0.6269} & \underline{0.6651} & 0.6658 & \underline{0.6533} & 0.6671 & \underline{0.6556} & \underline{0.6576}\\
        \midrule
        \midrule
        Ours & CAEN & \textbf{0.6789} & \textbf{0.6282} & \textbf{0.6801} & \textbf{0.5999} & \textbf{0.6790} & \textbf{0.6371} & \textbf{0.6746} & \textbf{0.6283} & \textbf{0.6786} & \textbf{0.6214}\\
        \bottomrule
    \end{tabular}
    }
    \label{tab:overall_perf}
\end{table*}

Different from these three methods, TIEN is the only one that gets equal or higher AUC than UB-GRUA. The result is rational since TIEN is the only one that does not just apply a symmetric structure in item behavior modeling compared to the user side, but tries to capture the item changes with a time-aware evolution model. Further, our proposed CAEN takes use of the field knowledge and models the item evolution over attribute changes, thus achieving a significantly higher AUC and lower Logloss than TIEN. This demonstrates that, 
it is more effective to model the item evolution under attribute granularity than that under interaction granularity and CAEN is the best fit for our fast-growing e-commerce environment. It is also worth noting that the performance improvement by introducing item behavior modeling (CAEN vs. UB-GRUA) is at the same level as that by introducing user behavior modeling (UB-GRUA vs. PNN). More discussions are expanded in the following analysis.



\subsection{Ablation Study (RQ2)}

To figure out the contribution of each component to the performance of our CAEN, three variants of CAEN are implemented for ablation experiments:
\begin{itemize}
    \item \textbf{CAEN-NS (SPL+HAN+FEL)} does not have the State Evolution Layer (SEL) in capturing the temporal dependencies of dynamic representation under chronological states, but replaces it with a fully connected layer.
    \item \textbf{CAEN-NH (SPL+TAL+SEL)} does not have the Attribute Attention Layer in Hierarchical Attention Network (AAL), but aggregates the corresponding users in each state by averaging pooling.
    \item \textbf{CAEN-NF (SPL+HAN+SEL)} does not have the Frequency Extraction Layer (FEL), thus ignores the activity level information of seller's operation.
\end{itemize}

\begin{table*}[t!]
    \caption{Performance of CAEN and its ablation models.}
    \centering
    \setlength{\tabcolsep}{1.5mm}{
    \begin{tabular}{c|cc|cc|cc|cc|cc}
        \toprule
        \multirow{2}{*}{Method}
        & \multicolumn{2}{c|} {overall} 
        & \multicolumn{2}{c|} {FMCG}
        & \multicolumn{2}{c|} {FA} 
        & \multicolumn{2}{c|} {GM}
        & \multicolumn{2}{c} {EL}\\ 
        & AUC & Logloss & AUC & Logloss & AUC & Logloss & AUC & Logloss & AUC & Logloss\\
        \midrule
        CAEN-NS & 0.6616 & 0.6498 & 0.6612 & 0.6204 & 0.6643 & 0.6552 & 0.6534 & 0.6586 & 0.6552 & 0.6516\\
        CAEN-NH & 0.6622 & 0.6472 & 0.6620 & 0.6189 & 0.6649 & 0.6528 & 0.6538 & 0.6550 & 0.6558 & 0.6473 \\
        CAEN-NF & 0.6622 & 0.6631 & 0.6616 & 0.6320 & 0.6650 & 0.6702 & 0.6536 & 0.6684 & 0.6556 & 0.6649 \\
        \midrule
        \textbf{CAEN} & \textbf{0.6789} & \textbf{0.6282} & \textbf{0.6801} & \textbf{0.5999} & \textbf{0.6790} & \textbf{0.6371} & \textbf{0.6746} & \textbf{0.6283} & \textbf{0.6786} & \textbf{0.6214} \\
        \bottomrule
    \end{tabular}
    }
    \label{tab:ablation}
\end{table*}

Note that Personalized Attention Layer (PAL) is reserved in all variants to obtain a personalized item representation. Table \ref{tab:ablation} presents the numerical results of ablation studies. By comparing the performances between CAEN and its variants, we observe that removing any component in CAEN will result in a performance decrease, which proves the effectiveness of each proposed layer. 
Firstly, the State Evolution Layer (SEL) is conducive to the item representation under each attribute by capturing the temporal dependency. This is because the item representation under each attribute is determined not only by the state of that time, but also by its past. Secondly, the attributes-aware attention mechanism in AAL can help to extract the interactions related to the item attribute, by putting more weight on users that are attracted by the attribute other than the users' constant interests, which also improves the performance significantly. Thirdly, the Frequency Extraction Layer (FEL) works like a ``supervisor of seller operation", which models the evolution over the timestamp of attribute modification and captures the seller's operational activities, is also proved to be effective.

\subsection{Practicability Study (RQ3)}

To verify the practicability of the proposed item behavior modeling method, we implement CAEN as a configurable module into different user behavior modeling methods, and compare the performance of individual GRU4Rec, ATRANK, DIEN, UB-GRUA to those with CAEN. Results in Table \ref{tab:practicability} prove the effectiveness and practicability of our methods. CAEN can be applied together with the state-of-the-art user sequential modeling methods and further improve their recommendation accuracy, especially for UB-GRUA.

\begin{table*}[t!]
    \caption{Performance of CAEN along with different user behavior sequential modeling.}
    \centering
    \setlength{\tabcolsep}{1.5mm}{
    \begin{tabular}{c|cc|cc|cc|cc|cc}
        \toprule
        \multirow{2}{*}{Method}
        & \multicolumn{2}{c|} {overall} 
        & \multicolumn{2}{c|} {FMCG}
        & \multicolumn{2}{c|} {FA} 
        & \multicolumn{2}{c|} {GM}
        & \multicolumn{2}{c} {EL}\\ 
        & AUC & Logloss & AUC & Logloss & AUC & Logloss & AUC & Logloss & AUC & Logloss\\
        \midrule
        GRU4Rec & 0.6579 & 0.6771 & 0.6565 & 0.6457 & 0.6607 & 0.6842 & 0.6500 & 0.6829 & 0.6511 & 0.6789 \\
        GRU4Rec + CAEN & 0.6589 & 0.6562 & 0.6576 & 0.6263 & 0.6617 & 0.6633 & 0.6509 & 0.6608 & 0.6536 & 0.6579 \\
        \midrule
        Caser & 0.6593 & 0.6708 & 0.6580 & 0.6415 & 0.6623 & 0.6768 & 0.6509 & 0.6779 & 0.6519 & 0.6730 \\
        Caser + CAEN & 0.6597 & 0.6771 & 0.6583 & 0.6440 & 0.6626 & 0.6828 & 0.6517 & 0.6881 & 0.6525 & 0.6782\\ 
        \midrule
        ATRANK & 0.6565 & 0.6567 & 0.6556 & 0.6271 & 0.6591 & 0.6630 & 0.6489 & 0.6625 & 0.6499 & 0.6614 \\
        ATRANK + CAEN & 0.6583 & 0.6549 & 0.6575 & 0.6274 & 0.6609 & 0.6615 & 0.6506 & 0.6582 & 0.6516 & 0.6583 \\
        \midrule
        DIEN & 0.6584 & 0.6606 & 0.6570 & 0.6320 & 0.6611 & 0.6679 & 0.6507 & 0.6644 & 0.6521 & 0.6588 \\
        DIEN + CAEN & 0.6598 & 0.6608 & 0.6584 & 0.6320 & 0.6627 & 0.6669 & 0.6515 & 0.6676 & 0.6529 & 0.6610 \\
        \midrule
        UB-GRUA & 0.6620 & 0.6567 & 0.6618 & 0.6310 & 0.6645 & 0.6623 & 0.6539 & 0.6623 & 0.6558 & 0.6564 \\
        UB-GRUA + CAEN & 0.6789 & 0.6282 & 0.6801 & 0.5999 & 0.6790 & 0.6371 & 0.6746 & 0.6283 & 0.6786 & 0.6214 \\
        \bottomrule
    \end{tabular}
    }
    \label{tab:practicability}
\end{table*}

\subsection{Performance on items with different levels of attribute dynamics (RQ4)}

\begin{figure*}[t!]
\centering
\includegraphics[width=11cm]{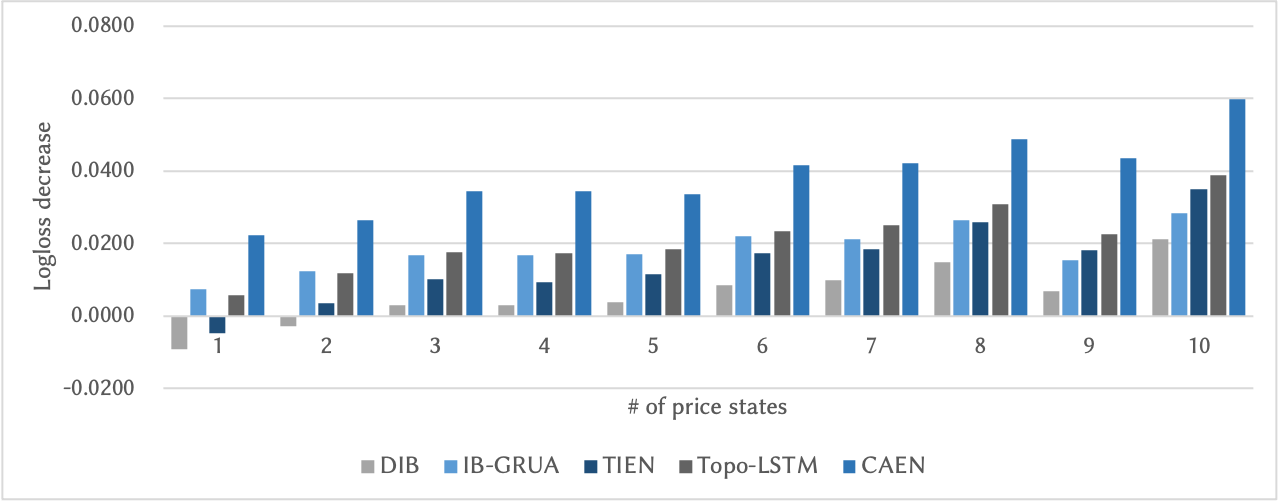}
\caption{Performance on items with a different number of price states. The vertical axis is the Logloss decrease compared to UB-GRUA.}
\label{fig:case_study}
\end{figure*}

To further analyze the role of our method in the recommendation system of fast-growing e-commerce, we evaluate the model performances over items with different level of attribute dynamics. To be more specific, we classify the samples by the number of price states of the target item, then measure their Logloss respectively. Here, we use Logloss because AUC is a metric of ordering relation, which is not appropriate for measuring the performance of a subset over the whole dataset. Results are shown in Figure \ref{fig:case_study}. As the number of price states increases from $1$ to $8$, the CTR prediction improvement caused by item behavior modeling increases under all methods. It happens because the items might be more popular (appear more times in the training dataset) while having more price states.
This also indicates the importance of modeling on item behaviors as the item changes increase. Compared with other item behavior modeling methods, CAEN can always benefit from a larger number of price states thanks to the SEL and PAL, which could share information within different price states of one specific item. When the number of price states is $9$, all other methods get a significant decrease than the number of price states as $8$, even lower than the number of price states as $7$, CAEN achieves better results than the number of price states as $7$, with only a bit decrease than the number of price states as $8$.

\subsection{Hyperparameter Analysis (RQ5)}

\begin{figure}[t!]
     \centering
     \begin{subfigure}[b]{0.45\textwidth}
         \centering
         \includegraphics[width=0.5\textwidth]{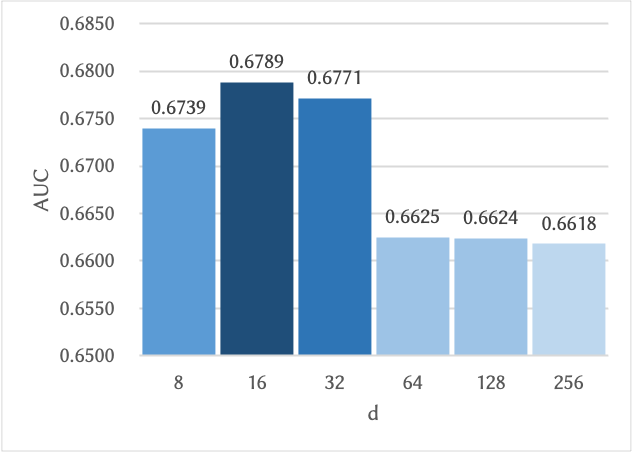}
         \caption{representation size}
         \label{fig:auc-size}
     \end{subfigure}
     \begin{subfigure}[b]{0.45\textwidth}
         \centering
         \includegraphics[width=0.5\textwidth]{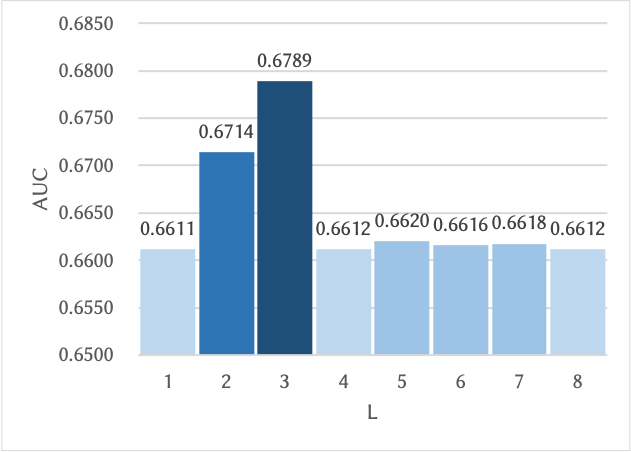}
         \caption{truncation length}
         \label{fig:auc-trunc}
     \end{subfigure}
    \caption{Performance with different hyperparameters.}
    \label{fig:hyperparameter}
\end{figure}

We evaluate the effect of two critical hyper-parameters: the size of item representation $d$ and the truncation length of item state $L$. Figure \ref{fig:auc-size} presents the AUC under the size of item representation $d$ in \{8, 16, 32, 64, 128, 256\}. As shown, the model performs better as $d$ increases from $8$ to $16$, but then worse as $d$ is further increased.
Figure \ref{fig:auc-trunc} presents the AUC under the truncation length in \{1,2,3,4,5,6,7,8\}. Increasing truncation length $L$ of the item state from $1$ to $3$ significantly improves model performance. However, when $L$ is larger or equal to $4$, the performance stays at a lower level which is similar to the performance when $L=1$. On one hand, considering more price states might also introduce noisy information. On the other hand, most of the samples are based on items that have a price change time less than or equal to $3$, thus it is enough to take the latest $3$ prices into account for them. 

\section{Conclusions and Future Work}
In this paper, we propose a novel method CAEN, for the modeling of item evolution over item attributes. To the best of our knowledge, we are the first to model item dynamics from the perspective of attribute change. The sequence of users who have interacted with the target item is partitioned according to the value of item attribute, and a hierarchical attention network is devised over it and works throughout the item evolution modeling. 
Results from the extensive experiments over actual e-commerce datasets in Southeast Asia show that our approach is superior to the state-of-art baselines, especially on the items with changes over attributes, which helps item recommendation to adapt to the growth of e-commerce.

This work still has some limitations. Firstly, studies have proposed CTR prediction schemes other than Equation (\ref{eq:ctr}). In the future, we will improve the downstream application of CAEN by regarding it as an enhanced item embedding method, thus it can be further used in more schemes. Secondly, recommended items may be affected by not only the manual operations on a single attribute, but also the manual operations on multiple attributes, and even the environmental factors. In the future, we will investigate how to capture the impact from multiple variables on user-item interactions, as well as the synergy among these variables.


\bibliographystyle{ACM-Reference-Format}
\bibliography{sample-base}










\end{document}